\documentclass[aps,prl,reprint]{revtex4-1}
\usepackage{amsfonts,amssymb,amsmath,bm,graphicx,color}
\usepackage[colorlinks=true,citecolor=blue,linkcolor=blue,urlcolor=blue]{hyperref}
\allowdisplaybreaks[4]
\newcommand\cc{{\rm c.c.}}
\DeclareMathOperator\tr{tr}
\begin{document}
%--- Front Matter
\title{Theory of orbital magnetic quadrupole moment and magnetoelectric susceptibility}
\author{Atsuo Shitade}
\affiliation{RIKEN Center for Emergent Matter Science, 2-1 Hirosawa, Wako, Saitama 351-0198, Japan}
\author{Hikaru Watanabe}
\affiliation{Department of Physics, Graduate School of Science, Kyoto University, Kyoto 606-8502, Japan}
\author{Youichi Yanase}
\affiliation{Department of Physics, Graduate School of Science, Kyoto University, Kyoto 606-8502, Japan}
\date{\today}
\begin{abstract}
  We derive a quantum-mechanical formula of the orbital magnetic quadrupole moment (MQM) in periodic systems by using the gauge-covariant gradient expansion.
  This formula is valid for insulators and metals at zero and nonzero temperature.
  We also prove a direct relation between the MQM and magnetoelectric (ME) susceptibility for insulators at zero temperature.
  It indicates that the MQM is a microscopic origin of the ME effect.
  Using the formula, we quantitatively estimate these quantities for room-temperature antiferromagnetic semiconductors BaMn$_2$As$_2$ and CeMn$_2$Ge$_{2 - x}$Si$_x$.
  We find that the orbital contribution to the ME susceptibility is comparable with or even dominant over the spin contribution.
\end{abstract}
\maketitle
%--- Main Matter
In classical electromagnetism, electric and magnetic multipole moments characterize the anisotropy of the charge and charge current densities.
Spin is also an important origin of the magnetic dipole moment.
When electric or magnetic dipole moments align in a certain direction, it is called a ferroelectrics or a ferromagnet.
In several decades, we have witnessed the importance of higher-order multipole moments in strongly correlated electron systems~\cite{RevModPhys.81.807,JPSJ.78.072001}.
More recently, higher-order topological insulators with electric multipole moments
were theoretically proposed~\cite{Benalcazar61,PhysRevB.96.245115,PhysRevLett.119.246401,PhysRevLett.119.246402}
and soon later realized in a metamaterial~\cite{Serra-Garcia2018} and a microwave circuit~\cite{Peterson2018}.
Multipole moments are now more ubiquitous than in the $19$th century when classical electromagnetism was established.

Among multipole moments, the magnetic quadrupole moment (MQM) has been believed to be an important ingredient
for the magnetoelectric (ME) effect~\cite{PhysRevB.76.214404,0953-8984-20-43-434203,PhysRevB.88.094429}.
In this phenomenon, the charge polarization is induced by a magnetic field, and the magnetization is induced by an electric field.
Both the inversion and time-reversal symmetries should be broken.
Although the multipole order observed in some $f$-electron systems~\cite{RevModPhys.81.807,JPSJ.78.072001} does not break the inversion symmetry,
the symmetry conditions are satisfied in the presence of the MQM.
Cr$_2$O$_3$ was the first material in which the ME effect was theoretically predicted~\cite{Dzyaloshinskii1960}
and experimentally observed~\cite{Astrov1960,Astrov1961,PhysRevLett.6.607,PhysRevLett.7.310}.
So far, the toroidal moment, which is the antisymmetric part of the MQM, was investigated
in Ga$_{2 - x}$Fe$_x$O$_3$~\cite{Popov1998,JPSJ.74.1419}, LiCoPO$_4$~\cite{nature06139,ncomms5796}, and Ni$_{0.4}$Mn$_{0.6}$TiO$_3$~\cite{ncomms3063},
and the symmetric MQM in Ba(TiO)Cu$_4$(PO$_4$)$_4$~\cite{ncomms13039}.
We also note that the ME effect has been intensively studied in the field of multiferroics
since the celebrated discoveries of BiFeO$_3$ thin films~\cite{Wang1719} and TbMnO$_3$~\cite{Kimura2003}.
Theoretically, an expression of the spin MQM was derived using an adiabatic deformation~\cite{PhysRevLett.101.077203,PhysRevB.93.195167}.
However, it is not gauge invariant, nor does it take spin-orbit interactions into account,
and another expression was thermodynamically derived using the semiclassical theory~\cite{PhysRevB.97.134423}.

Although the above materials are magnetic insulators, the MQM also appears in electron systems.
In a zigzag chain~\cite{JPSJ.83.014703,JPSJ.84.064717} and a honeycomb lattice~\cite{PhysRevB.90.024432,PhysRevB.90.081115,0953-8984-28-39-395601},
the inversion symmetry may be broken by a magnetic order, leading to the spin MQM and ME effect.
It was pointed out that a spin magnetic hexadecapole moment appears in an antiferromagnetic (AFM) semiconductor BaMn$_2$As$_2$~\cite{PhysRevB.96.064432}.
In these theoretical studies, the orbital contribution has been neglected although it may not be negligible.

When we calculate the electric or magnetic multipole moments quantum mechanically in periodic systems, we suffer from the fact that the position operator is unbounded.
This difficulty can be solved in several ways.
The charge polarization $P^i$ was formulated
by calculating the charge current density $J^i$ induced by an adiabatic deformation of the Hamiltonian~\cite{PhysRevB.47.1651,PhysRevB.48.4442,RevModPhys.66.899}.
This idea relies on the electromagnetic relation $J^i = {\dot P}^i + \epsilon^{ijk} \partial_{X^j} M_k$,
in which $\epsilon^{ijk}$ is the totally antisymmetric tensor.
The result is expressed by the Berry connection and valid only for insulators at zero temperature.
Later, it was reformulated using the Green's function~\cite{PhysRevB.84.205137,PhysRevB.88.155121,JPSJ.83.033708}.
On the other hand, the orbital magnetization $M_k$ was defined by the thermodynamic relation $M_k \equiv -\partial \Omega/\partial B^k$~\cite{PhysRevLett.99.197202},
in which $\Omega$ is the grand potential, and $B^k$ is a magnetic field.
The result is expressed by the Berry curvature and magnetic moment and is valid for insulators, without or with the Chern number, and metals at zero and nonzero temperature.
Effects of disorder and interactions were studied with the help of the Green's function~\cite{PhysRevB.84.205137,PhysRevB.86.214415,PhysRevB.90.125132}.

In this Rapid Communication, we derive a quantum-mechanical formula of the orbital MQM in periodic systems.
First, we define the MQM and prove a direct relation to the ME susceptibility based on thermodynamic relations.
This relation indicates that the MQM is a microscopic origin of the ME effect.
Next, we calculate the orbital MQM in the Bloch basis using the gauge-covariant gradient expansion of the Keldysh Green's function~\cite{0953-8984-6-39-010,LEVANDA2001199,1708.03424}.
Finally, we apply these results to the AFM semiconductors BaMn$_2$As$_2$ and CeMn$_2$Ge$_{2 - x}$Si$_x$.
We find that the orbital contribution to the ME susceptibility is comparable with or even dominant over the spin contribution.

We begin with the thermodynamic relation of the grand potential $\Omega \equiv E - T S - \mu N$,
\begin{equation}
  {\rm d} \Omega
  = -S {\rm d} T - M_k {\rm d} B^k  - N {\rm d} \mu, \label{eq:maxwell1}
\end{equation}
in which $S, N$ are the entropy and particle number, and $T, \mu$ are temperature and the chemical potential.
Supposing a magnetic field $B(X)$ is nonuniform and varies slowly compared with a length scale of the lattice constants,
then we can extend Eq.~\eqref{eq:maxwell1} to a local relation,
\begin{equation}
  {\rm d} \Omega = -S {\rm d} T - (M_k - \partial_{X^l} M^l_{\phantom{l} k}) {\rm d} B^k - N {\rm d} \mu. \label{eq:maxwell2}
\end{equation}
$M^l_{\phantom{l} k}$ is the MQM and in general not symmetric over $l$ and $k$.
The magnetic toroidal and monopole moments are also included in the $3 \times 3 = 9$ components of $M^l_{\phantom{l} k}$.
By integrating by parts, we obtain a general definition of the MQM,
\begin{equation}
  M^l_{\phantom{l} k}
  \equiv -\frac{\partial \Omega}{\partial (\partial_{X^l} B^k)}, \label{eq:maxwell3}
\end{equation}
together with the well-known relations $S = -\partial \Omega/\partial T$ and $N = -\partial \Omega/\partial \mu$.
We also obtain the Maxwell relations,
\begin{subequations} \begin{align}
  -\frac{\partial^2 \Omega}{\partial T \partial (\partial_{X^l} B^k)}
  = & \frac{\partial S}{\partial (\partial_{X^l} B^k)}
  = \frac{\partial M^l_{\phantom{l} k}}{\partial T}, \label{eq:maxwell4a} \\
  -\frac{\partial^2 \Omega}{\partial (\partial_{X^l} B^k) \partial \mu}
  = & \frac{\partial M^l_{\phantom{l} k}}{\partial \mu}
  = \frac{\partial N}{\partial (\partial_{X^l} B^k)}. \label{eq:maxwell4b}
\end{align} \label{eq:maxwell4}\end{subequations}

The first relation~\eqref{eq:maxwell4a} is practically important.
To see this, we define a related quantity,
\begin{equation}
  {\tilde M}^l_{\phantom{l} k}
  \equiv -\frac{\partial K}{\partial (\partial_{X^l} B^k)}, \label{eq:maxwell5}
\end{equation}
which involves the energy $K \equiv E - \mu N = \Omega + T S$.
Using Eq.~\eqref{eq:maxwell4a}, these two are related by
\begin{align}
  {\tilde M}^l_{\phantom{l} k}
  = & -\frac{\partial \Omega}{\partial (\partial_{X^l} B^k)} - T \frac{\partial S}{\partial (\partial_{X^l} B^k)} \notag \\
  = & M^l_{\phantom{l} k} - T \frac{\partial M^l_{\phantom{l} k}}{\partial T}
  = \frac{\partial (\beta M^l_{\phantom{l} k})}{\partial \beta}. \label{eq:maxwell6}
\end{align}
We calculate Eq.~\eqref{eq:maxwell5} and solve Eq.~\eqref{eq:maxwell6} to obtain the MQM.
A similar relation is known for the orbital magnetization~\cite{PhysRevLett.99.197202}.

A direct relation between the MQM and ME susceptibility follows from the second relation~\eqref{eq:maxwell4b}.
When the system is an insulator at zero temperature,
the charge density can be expressed by $q N = - \partial_{X^i} P^i$, with $q$ being the electron charge,
and Eq.~\eqref{eq:maxwell4b} is reduced to
\begin{equation}
  -q \frac{\partial M^l_{\phantom{l} k}}{\partial \mu}
  = \frac{\partial (\partial_{X^i} P^i)}{\partial (\partial_{X^l} B^k)}
  = \alpha^l_{\phantom{l} k}. \label{eq:maxwell7}
\end{equation}
$\alpha^l_{\phantom{l} k} \equiv \partial P^l/\partial B^k$ is the linear ME susceptibility.
If the system is a metal or at nonzero temperature,
the polarization charge is screened by the itinerant or thermally excited charge, and hence this relation does not make sense.
This relation is valid for the orbital and spin contributions and indicates that the MQM is a microscopic origin of the ME effect.
Gao {\it et al.} obtained the same relation but restricted their discussion to the spin toroidal moment~\cite{PhysRevB.97.134423}.

Let us comment on the St\v{r}eda formula for the MQM.
The St\v{r}eda formula relates the Hall conductivity to the orbital magnetization
as $\partial J^i/\partial E_j = q \epsilon^{ijk} \partial M_k/\partial \mu = q \epsilon^{ijk} \partial N/\partial B^k$~\cite{0022-3719-15-22-005}.
The first equality is explained by the magnetization current $J^i = \epsilon^{ijk} \partial_{X^j} M_k = \epsilon^{ijk} (\partial_{X^j} \mu) (\partial M_k/\partial \mu)$
and identifying $\partial_{X^j} \mu/q$ as an electric field $E_j$.
The second equality follows from the Maxwell relation.
Similarly, the magnetization is expressed by $M_k = -\partial_{X^l} M^l_{\phantom{l} k} = -(\partial_{X^l} \mu) (\partial M^l_{\phantom{l} k}/\partial \mu)$,
leading to $\partial M_k/\partial E_l = -q \partial M^l_{\phantom{l} k}/\partial \mu$.
This electric-field-induced magnetization is defined in insulators and metals at zero and nonzero temperature.
However, in the above identification, we do not take into account the dissipation effect caused by the electric field on the Fermi surface.
The spin~\cite{Edelstein1990233,Aronov1989} and orbital Edelstein effects~\cite{Yoda2015,PhysRevLett.116.077201} are known as such Fermi-surface terms.
Therefore, the St\v{r}eda formulas are valid only for insulators at zero temperature.
Combining Eq.~\eqref{eq:maxwell7}, we obtain 
\begin{equation}
  \partial M_k/\partial E_l
  = -q \partial M^l_{\phantom{l} k}/\partial \mu
  = \partial P^l/\partial B^k. \label{eq:streda}
\end{equation}
This is not trivial because the charge polarization is not a thermodynamic quantity but a geometric one,
while the magnetization is a thermodynamic one.
Note that the above discussion holds for disordered and interacting systems because it relies on thermodynamics.
Below, we microscopically prove Eq.~\eqref{eq:streda} for the orbital contribution in the absence of disorder or interactions.

Here, we derive the quantum-mechanical formula of the orbital MQM.
To calculate the energy in the nonuniform magnetic field, we use the gauge-covariant gradient expansion~\cite{0953-8984-6-39-010,LEVANDA2001199,1708.03424}.
In this method, we attach the Wilson line to the Keldysh Green's function, which guarantees the gauge covariance,
and express the gauge-covariant Keldysh Green's function in terms of the center-of-mass coordinate $X$ and the relative momentum $p$.
As a result, the convolution in the Dyson equation turns into the noncommutative Moyal product. 
In the absence of disorder or interactions, the variation of the energy due to the nonuniform magnetic field is given by~\cite{suppl}
\begin{widetext}
\begin{align}
  K_{D {\cal F}}
  = & -\frac{i \hbar^2}{6} \partial_{X^l} {\cal F}_{ij} \int \frac{{\rm d}^d p}{(2 \pi \hbar)^d} \int \frac{{\rm d} \xi}{2 \pi} f(\xi) \xi \notag \\
  & \times \tr [g_0^{\rm R} \partial_{p_l} (g_0^{\rm R})^{-1} g_0^{\rm R} \partial_{p_i} (g_0^{\rm R})^{-1} g_0^{\rm R} \partial_{p_j} (g_0^{\rm R})^{-1} g_0^{\rm R}
  + g_0^{\rm R} \partial_{p_j} (g_0^{\rm R})^{-1} g_0^{\rm R} \partial_{p_i} (g_0^{\rm R})^{-1} g_0^{\rm R} \partial_{p_l} (g_0^{\rm R})^{-1} g_0^{\rm R}] + \cc, \label{eq:omq1}
\end{align}
in which ${\cal F}_{ij} = q \epsilon_{ijk} B^k$ is the magnetic field,
$d$ is the space dimension,
and $g_0^{\rm R}(\xi, {\vec p}) = [\xi - {\cal H}({\vec p}) + \mu + i \eta]^{-1}$ $(\eta \to +0)$ is the retarded Green's function of the Hamiltonian ${\cal H}({\vec p})$.
By expanding the trace in Eq.~\eqref{eq:omq1} with respect to the Bloch basis that satisfies ${\cal H}({\vec p}) | u_n({\vec p}) \rangle = \epsilon_n({\vec p}) | u_n({\vec p}) \rangle$,
we obtain
\begin{subequations} \begin{align}
  {\tilde M}^l_{\phantom{l} k}
  = & \frac{q}{\hbar} \frac{1}{2} \epsilon_{ijk} \sum_n \int \frac{{\rm d}^d p}{(2 \pi \hbar)^d}
  \{A_n^{lij} f_n (\epsilon_n - \mu) + m_n^{lij} [f_n + f_n^{\prime} (\epsilon_n - \mu)] + \gamma_n^{lij} [2 f_n^{\prime} + f_n^{\prime \prime} (\epsilon_n - \mu)]\}, \label{eq:omq2a} \\
  A_n^{lij}
  \equiv & \frac{\hbar^3}{2} \sum_{m, r (\not= n)} \frac{\langle u_n | v^l | u_m \rangle \langle u_m | v^i | u_r \rangle \langle u_r | v^j | u_n \rangle}{(\epsilon_n - \epsilon_m)^2 (\epsilon_n - \epsilon_r)}
  + \frac{\hbar^3}{2} \sum_{m (\not= n)} \frac{\langle u_n | v^l | u_m \rangle \langle u_m | v^j | u_n \rangle \langle u_n | v^i | u_n \rangle}{(\epsilon_n - \epsilon_m)^3}
  + \cc - (i \leftrightarrow j) \notag \\
  = & \frac{\hbar^3}{2} \sum_{m (\not= n)} \frac{\langle \partial_{p_l} u_n | u_m \rangle \langle u_m | \partial_{p_i} (\epsilon_n + {\cal H}) Q_n | \partial_{p_j} u_n \rangle}{\epsilon_n - \epsilon_m}
  + \cc - (i \leftrightarrow j), \label{eq:omq2b} \\
  m_n^{lij}
  \equiv & -\frac{\hbar^3}{6} \sum_{m, r (\not= n)} \frac{\langle u_n | v^l | u_m \rangle \langle u_m | v^i | u_r \rangle \langle u_r | v^j | u_n \rangle}{(\epsilon_n - \epsilon_m) (\epsilon_n - \epsilon_r)}
  - \frac{\hbar^3}{3} \sum_{m (\not= n)} \frac{\langle u_n | v^l | u_m \rangle \langle u_m | v^j | u_n \rangle \langle u_n | v^i | u_n \rangle}{(\epsilon_n - \epsilon_m)^2}
  + \cc - (i \leftrightarrow j) \notag \\
  = & -\frac{\hbar^3}{6} \langle \partial_{p_l} u_n | Q_n \partial_{p_i} (2 \epsilon_n + {\cal H}) Q_n | \partial_{p_j} u_n \rangle
  + \cc - (i \leftrightarrow j), \label{eq:omq2c} \\
  \gamma_n^{lij}
  \equiv & \frac{\hbar^3}{12} \sum_{m (\not= n)} \frac{\langle u_n | v^l | u_m \rangle \langle u_m | v^j | u_n \rangle \langle u_n | v^i | u_n \rangle}{\epsilon_n - \epsilon_m}
  + \cc - (i \leftrightarrow j) \notag \\
  = & \frac{\hbar^3}{12} \langle \partial_{p_l} u_n | (\epsilon_n - {\cal H}) | \partial_{p_j} u_n \rangle \partial_{p_i} \epsilon_n
  + \cc - (i \leftrightarrow j). \label{eq:omq2d}
\end{align} \label{eq:omq2}\end{subequations}
\end{widetext}
Here, $v^i \equiv \partial_{p_i} {\cal H}$ is the velocity operator, $f_n \equiv f(\epsilon_n - \mu)$ is the Fermi distribution function,
and $Q_n \equiv 1 - | u_n \rangle \langle u_n |$ is the projection operator, which guarantees the gauge invariance.
For degenerate bands, we have to modify the projection operator as $Q_n \equiv 1 - \sum_s | u_{ns} \rangle \langle u_{ns} |$,
in which $s$ indicates the index of degenerate bands with energy $\epsilon_n$.
The arguments $\xi, {\vec p}$ are dropped for simplicity.
By solving Eq.~\eqref{eq:maxwell6}, we obtain our central result on the orbital MQM,
\begin{align}
  M^l_{\phantom{l} k}
  = & \frac{q}{\hbar} \frac{1}{2} \epsilon_{ijk} \sum_n \int \frac{{\rm d}^d p}{(2 \pi \hbar)^d} \notag \\
  & \times \left[-A_n^{lij} \int_{\epsilon_n - \mu}^{\infty} {\rm d} z f(z) + m_n^{lij} f_n + \gamma_n^{lij} f_n^{\prime}\right]. \label{eq:omq3}
\end{align}
The third term seems to be a Fermi-surface term unlike thermodynamic quantities.
However, it can be integrated by parts because $\gamma_n^{lij}$ is proportional to $\partial_{p_i} \epsilon_n$.
Therefore, the second and third terms are Fermi-sea terms.

For insulators at zero temperature, where we can drop the derivatives of the Fermi distribution function, we obtain
\begin{equation}
  -q \frac{\partial M^l_{\phantom{l} k}}{\partial \mu}
  = \frac{q^2}{\hbar} \frac{1}{2} \epsilon_{ijk} \sum_n^{\rm occ} \int \frac{{\rm d}^d p}{(2 \pi \hbar)^d} A_n^{lij}
  = \alpha^l_{\phantom{l} k}. \label{eq:omq4}
\end{equation}
This formula is identical to the orbital ME susceptibility derived by using an adiabatic deformation in the context of topological insulators~\cite{PhysRevLett.102.146805,PhysRevB.81.205104},
except for the isotropic Chern-Simons $3$-form.
Our formula is gauge invariant and hence does not include such a gauge-dependent term.
Thus, we have microscopically proved Eq.~\eqref{eq:streda} for the orbital contribution.
The full ME susceptibility of Cr$_2$O$_3$ including spin and lattice~\cite{PhysRevB.86.094430} was calculated by first principles.
Our formula of the orbital MQM Eq.~\eqref{eq:omq3} is based on the Bloch basis, and thus it enables a first-principles calculation of the orbital MQM.

It is suggestive to compare Eq.~\eqref{eq:omq3} with the quantum-mechanical formula of the orbital magnetization~\cite{PhysRevLett.99.197202}
that consists of the Berry-curvature and magnetic-moment terms.
These are interpreted as magnetizations arising from the itinerant and local circulations, respectively,
in the semiclassical~\cite{PhysRevLett.95.137204,PhysRevLett.97.026603} and Wannier-basis theories~\cite{PhysRevLett.95.137205,PhysRevB.74.024408}.
Similarly, the first term in Eq.~\eqref{eq:omq3} is the itinerant contribution to the orbital MQM, while the second and third terms are the local ones.
In fact, according to Eq.~(5) in Ref.~\cite{PhysRevB.81.205104}, $A_n^{lij}$ is rewritten by a virtual process from an occupied band $n$ to an unoccupied band $m$
via the electric dipole moment ${\vec r}^{\prime}$ and magnetic dipole moment ${\vec r}^{\prime} \times {\vec v}$
and is consistent with the group-theoretical analysis.
Such an interband process is allowed not only in metals but also in insulators.
Therefore, the itinerant contribution is important even in insulators, as demonstrated below.

Let us apply our formula to real materials.
First, we focus on an AFM semiconductor BaMn$_2$As$_2$~\cite{PhysRevB.79.075120,PhysRevB.79.094519,PhysRevB.80.100403}.
In this material, since two Mn sites are crystallographically inequivalent even in the paramagnetic phase,
the AFM order breaks the time-reversal and inversion symmetries instead of the translation symmetry.
By hole doping, Ba$_{1 - x}$K$_x$Mn$_2$As$_2$ becomes a metal, but the AFM order is robust up to $x < 0.16$~\cite{PhysRevLett.108.087005,PhysRevB.87.144418}.
The group-theoretical analysis and microscopic calculation revealed that
this seemingly conventional AFM order is in fact the ferroic order of the magnetic hexadecapole moment and MQM~\cite{PhysRevB.96.064432}.
The ferroic MQM suggests a room-temperature ME effect below the N\'eel temperature $T_{\rm N} = 625~{\rm K}$.

We use an effective model of Mn $3 d_{x^2 - y^2}$ orbitals~\cite{PhysRevB.96.064432},
\begin{align*}
  {\cal H}({\vec q})
  = & \epsilon({\vec q}) + V({\vec q}) \rho^x + [{\vec g}({\vec q}) - {\vec h}] \cdot \rho^z {\vec \sigma}, \\
  \epsilon({\vec q})
  = & -2 t_1 (\cos q_x + \cos q_y) \notag \\
  & - 8 t_2 \cos q_x/2 \cos q_y/2 \cos q_z/2, \\
  V({\vec q})
  = & -4 v_1 \cos q_x/2 \cos q_y/2 - 2 v_2 \cos q_z/2, \\
  {\vec g}({\vec q})
  = &
  \begin{bmatrix}
    2 \alpha_1 \sin q_y + 8 \alpha_2 \cos q_x/2 \sin q_y/2 \cos q_z/2 \\
    2 \alpha_1 \sin q_x + 8 \alpha_2 \sin q_x/2 \cos q_y/2 \cos q_z/2 \\
    8 \alpha_3 \sin q_x/2 \sin q_y/2 \sin q_z/2
  \end{bmatrix},
\end{align*}
in which ${\vec \rho}, {\vec \sigma}$ are the Pauli matrices for the sublattice and spin degrees of freedom.
$q_x = k_x a, q_y = k_y a, q_z = k_z c$ are the dimensionless wave numbers with $a, c$ being the lattice constants.
$t_1, t_2$ and $v_1, v_2$ are the intra- and intersublattice transfer integrals, respectively,
${\vec g}({\vec q})$ represents the spin-orbit interaction, and ${\vec h}$ is the AFM mean field.
This model correctly captures the low-energy physics of this material.
First-principles calculation is needed for a quantitative prediction but is a future problem.

\begin{figure*}
  \centering
  \includegraphics[clip,width=\textwidth]{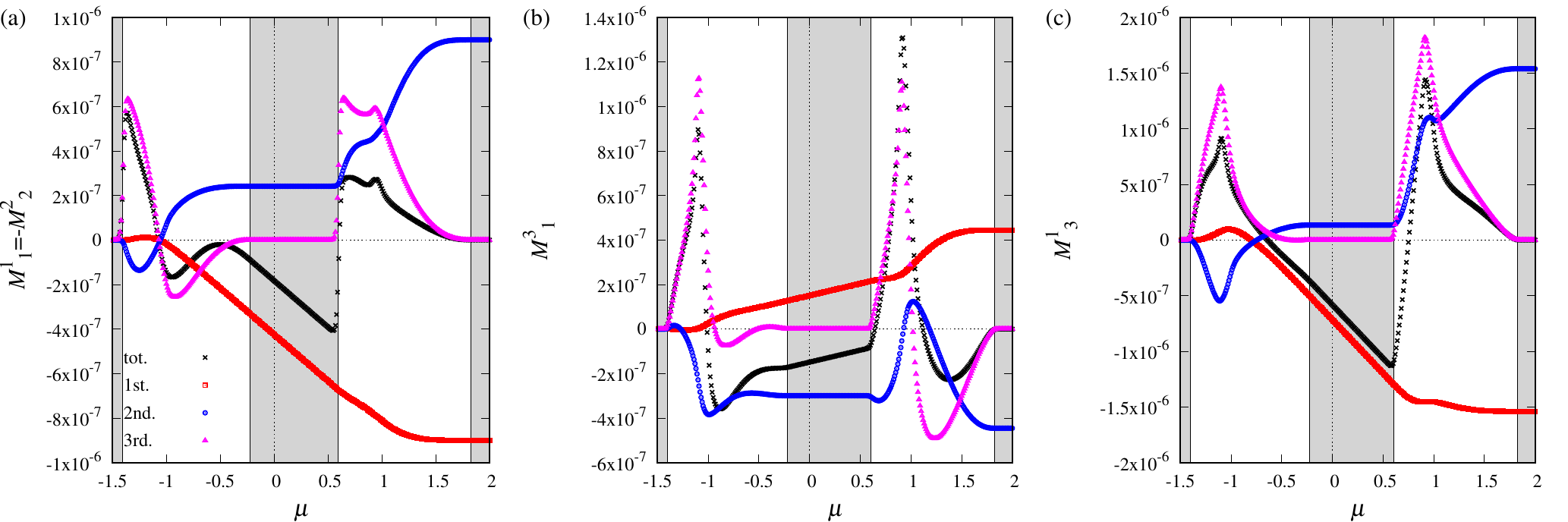}
  \caption{%
  Chemical potential dependence of the nonzero orbital MQM
  (a) for ${\vec h} = h {\vec z}$
  and (b), (c) for ${\vec h} = h {\vec x}$
  in the unit of $q h/\hbar$.
  The black star shows the total orbital MQM,
  the red square shows the itinerant contribution given by the first term in Eq.~\eqref{eq:omq3},
  and the blue circle and magenta triangle show the local contributions given by the second and third terms.
  The gray region shows that the system is an insulator.%
  } \label{fig:omq}
\end{figure*}

Figure~\ref{fig:omq}(a) shows the chemical potential dependence of the nonzero orbital MQM for ${\vec h} = h {\vec z}$.
Only $M^1_{\phantom{1} 1} = -M^2_{\phantom{2} 2}$ is allowed by the symmetry.
We use the same parameters as in Ref.~\cite{PhysRevB.96.064432}, i.e.,
$t_1 = -0.1, t_2 = -0.05, v_1 = 0.05, v_2 = 0.01, 2 \alpha_1 = -0.005, 8 \alpha_2 = 0.001, 8 \alpha_3 = 0.01, T = 0.01$
in the unit of $h = 1$.
The system size is given by $L_1 = L_2 = L_3 = 200$, and the lattice constants $a = 4.15~{\rm \AA}, c = 13.4~{\rm \AA}$ are taken from the experimental data~\cite{PhysRevB.87.144418}.
The itinerant contribution that has been neglected in the atomic~\cite{JPSJ.77.064710} or cluster~\cite{PhysRevB.95.094406} analysis of multipole moments is comparable with the local contributions.
If we turn off the spin-orbit interactions, namely, put $\alpha_1, \alpha_2, \alpha_3 = 0$, all the components vanish.
According to Eq.~\eqref{eq:streda}, the orbital MQM $M^l_{\phantom{l} k}$ linearly depends on the chemical potential $\mu$ when the system is an insulator,
as found for the spin toroidal moment~\cite{PhysRevB.97.134423},
and its slope with the minus sign is equal to the orbital ME susceptibility $\alpha^l_{\phantom{l} k}$.
The obtained orbital ME susceptibility is $\alpha^1_{\phantom{1} 1} = -\alpha^2_{\phantom{2} 2} = 4.2 \times 10^{-7} q^2/\hbar = 1.3 \times 10^{-4}~{\rm ps/m}$.
This is not negligible to the spin contribution $-1.5 \times 10^{-3}~{\rm ps/m}$,
which was estimated by using Eq.~(45) in Ref.~\cite{PhysRevB.96.064432} with $h = 1~{\rm eV}$ and the $g$ factor $g = 2$.

Figures~\ref{fig:omq}(b) and \ref{fig:omq}(c) show the same but for ${\vec h} = h {\vec x}$, in which $M^3_{\phantom{3} 1}, M^1_{\phantom{1} 3}$ are allowed.
Such a situation is indeed realized in another compound CeMn$_2$Ge$_{2 - x}$Si$_x$~\cite{MdDin2015}, which is isostructural with BaMn$_2$As$_2$.
Although the microscopic parameters for CeMn$_2$Ge$_{2 - x}$Si$_x$ should differ from those for BaMn$_2$As$_2$, we use the same parameters for comparison.
Regarding the former component of the ME susceptibility,
the orbital contribution $\alpha^3_{\phantom{3} 1} = -1.1 \times 10^{-7} q^2/\hbar = -3.3 \times 10^{-5}~{\rm ps/m}$ is dominant over the spin contribution $1.9 \times 10^{-6}~{\rm ps/m}$.
These are suppressed because the system is quasi-two-dimensional, i.e., $|t_1| > |t_2|$ and $|v_1| > |v_2|$.
Note that, in Cr$_2$O$_3$, the orbital contribution to the longitudinal component is ten times larger than the spin contribution~\cite{PhysRevB.86.094430}.
We also obtain the orbital contribution $\alpha^1_{\phantom{1} 3} = 9.6 \times 10^{-7} q^2/\hbar = 2.9 \times 10^{-4}~{\rm ps/m}$ and the spin contribution $1.5 \times 10^{-3}~{\rm ps/m}$.

To summarize, we have derived a quantum-mechanical formula of the orbital MQM in periodic systems by using the gauge-covariant gradient expansion of the Keldysh Green's function.
Based on the thermodynamic argument, we have defined the MQM as a response to a nonuniform magnetic field and proved a direct relation between the MQM and the ME susceptibility.
This relation indicates that the MQM is a microscopic origin of the ME effect.
The obtained formula of the orbital MQM is consistent with the orbital ME susceptibility in the literature.
We have applied the formula to an effective model of locally noncentrosymmetric AFM semiconductors BaMn$_2$As$_2$ and CeMn$_2$Ge$_{2 - x}$Si$_x$
and found that the itinerant contribution is comparable with the local contributions,
although the former has been neglected in the local analysis of multipole moments.
We have also found that the orbital contribution to the ME susceptibility is comparable with or dominant over the spin contribution.
According to the group-theoretical analysis, there are also many other materials with the MQM and high N\'eel temperature.
Our quantum-mechanical formula of the MQM can be implemented in first-principles calculations and provides a guideline for searching the large high-temperature ME effect.

{\it Note added.} Recently, we became aware of a related paper by Gao and Xiao~\cite{1803.06726}.
Their results on the orbital MQM, derived by the semiclassical theory, agree with ours.
They discuss a role of the orbital MQM in nonlinear anomalous thermoelectric transport.

%--- Acknowledgments
\begin{acknowledgments}
  A.S. thanks N.~Nagaosa for discussions on the St\v{r}eda formula and A.~Daido for pointing out that Eq.~\eqref{eq:omq4} does not include the Chern-Simons $3$-form.
  This work was supported by Grants-in-Aid for Scientific Research on Innovative Areas ``J-Physics'' (Grant No.~JP15H05884)
  and ``Topological Materials Science'' (Grant No.~JP16H00991) from the Japan Society for the Promotion of Science (JSPS),
  and by JSPS KAKENHI (Grants No.~JP15K05164 and No.~JP15H05745).
  A.S. was supported by the RIKEN Special Postdoctoral Researcher Program.
\end{acknowledgments}
%--- References
\end{document}

%--- Front Matter
\title{Supplemental Material for ``Theory of Orbital Magnetic Quadrupole Moment and Magnetoelectric Susceptibility''}
\author{Atsuo Shitade}
\affiliation{RIKEN Center for Emergent Matter Science, 2-1 Hirosawa, Wako, Saitama 351-0198, Japan}
\author{Hikaru Watanabe}
\affiliation{Department of Physics, Graduate School of Science, Kyoto University, Kyoto 606-8502, Japan}
\author{Youichi Yanase}
\affiliation{Department of Physics, Graduate School of Science, Kyoto University, Kyoto 606-8502, Japan}
\date{\today}
\maketitle
%--- Main Matter
\section{Gauge-Covariant Gradient Expansion of Keldysh Green's function} \label{sec:gradient}
Gauge-covariant gradient expansion is an efficient method to derive a perturbation theory of the Keldysh Green's function with respect to a gradient or a field strength.
In this method, the Keldysh Green's function ${\hat G}(x_1, x_2)$ is expressed in the Wigner representation,
namely, in terms of the center-of-mass coordinate $X_{12} \equiv (x_1 + x_2)/2$ and the relative momentum $p_{12}$.
To guarantee the gauge covariance, we should attach the Wilson line ${\hat W}(x_1, x_2)$ to the Keldysh Green's function,
${\tilde G}(x_1, x_2) \equiv {\hat W}(X_{12}, x_1) {\hat G}(x_1, x_2) {\hat W}(x_2, X_{12})$, before the Wigner transformation.
Since the Wigner transformation is the Fourier transformation with respect to the relative coordinate $x_{12} \equiv x_1 - x_2$,
convolution turns into a sort of product.
This product is noncommutative and called the Moyal product.
Although the Moyal product was obtained up to the infinite order with respect to a gradient and a U($1$) field strength,
we show only four terms necessary to derive a perturbation theory with respect to a magnetic-field gradient $\partial_{X^l} B^k$,
\begin{subequations} \begin{align}
  {\tilde A} \star {\tilde B}
  = & {\tilde A} {\tilde B} + (i \hbar/2) {\cal P}_D({\tilde A}, {\tilde B}) + (i \hbar/2) {\cal P}_{\cal F}({\tilde A}, {\tilde B})
  + (1/3) (i \hbar/2)^2 {\cal P}_{D {\cal F}}({\tilde A}, {\tilde B}), \label{eq:moyal1a} \\
  {\cal P}_D({\tilde A}, {\tilde B})
  = & \partial_{X^{\lambda}} {\tilde A} \partial_{p_{\lambda}} {\tilde B} - \partial_{p_{\lambda}} {\tilde A} \partial_{X^{\lambda}} {\tilde B}, \label{eq:moyal1b} \\
  {\cal P}_{\cal F}({\tilde A}, {\tilde B})
  = & {\cal F}_{\mu \nu} \partial_{p_{\mu}} {\tilde A} \partial_{p_{\nu}} {\tilde B}, \label{eq:moyal1c} \\
  {\cal P}_{D {\cal F}}({\tilde A}, {\tilde B})
  = & \partial_{X^{\lambda}} {\cal F}_{\mu \nu} (\partial_{p_{\mu}} {\tilde A} \partial_{p_{\lambda}} \partial_{p_{\nu}} {\tilde B}
  - \partial_{p_{\lambda}} \partial_{p_{\mu}} {\tilde A} \partial_{p_{\nu}} {\tilde B}). \label{eq:moyal1d}
\end{align} \label{eq:moyal1}\end{subequations}
The arguments $X, p$ are dropped for simplicity.
The first term indicates that convolution turns into the ordinary product in the presence of the translation symmetry as is well known.
${\cal F}_{j0} = q E_j, {\cal F}_{ij} = q \epsilon_{ijk} B^k$ are electromagnetic fields with $\epsilon_{ijk}$ being the totally antisymmetric tensor.

The Keldysh Green's function is determined by the Dyson equation,
\begin{equation}
  ({\tc L} - {\tilde \Sigma}) \star {\tilde G}
  = {\tilde G} \star ({\tc L} - {\tilde \Sigma})
  = 1, \label{eq:dyson1}
\end{equation}
in which ${\tc L}, {\tilde \Sigma}$ are the Lagrangian and self-energy.
Similarly to Eq.~\eqref{eq:moyal1a}, the Keldysh Green's function and self-energy are expanded as
\begin{subequations} \begin{align}
  {\tilde G}
  = & {\tilde G}_0 + (\hbar/2) {\tilde G}_D + (\hbar/2) {\tilde G}_{\cal F} + (1/3) (\hbar/2)^2 {\tilde G}_{D {\cal F}}, \label{eq:moyal2a} \\
  {\tilde \Sigma}
  = & {\tilde \Sigma}_0 + (\hbar/2) {\tilde \Sigma}_D + (\hbar/2) {\tilde \Sigma}_{\cal F} + (1/3) (\hbar/2)^2 {\tilde \Sigma}_{D {\cal F}}. \label{eq:moyal2b}
\end{align} \label{eq:moyal2}\end{subequations}
We substitute Eqs.~\eqref{eq:moyal1a} and \eqref{eq:moyal2} into the left Dyson equation to obtain
\begin{subequations} \begin{align}
  {\tilde G}_0
  = & ({\tc L} - {\tilde \Sigma}_0)^{-1}, \label{eq:dyson2a} \\
  {\tilde G}_0^{-1} {\tilde G}_D
  = & {\tilde \Sigma}_D {\tilde G}_0
  - i (\partial_{X^{\lambda}} {\tilde G}_0^{-1} \partial_{p_{\lambda}} {\tilde G}_0 - \partial_{p_{\lambda}} {\tilde G}_0^{-1} \partial_{X^{\lambda}} {\tilde G}_0), \label{eq:dyson2b} \\
  {\tilde G}_0^{-1} {\tilde G}_{\cal F}
  = & {\tilde \Sigma}_{\cal F} {\tilde G}_0
  - i {\cal F}_{\mu \nu} \partial_{p_{\mu}} {\tilde G}_0^{-1} \partial_{p_{\nu}} {\tilde G}_0, \label{eq:dyson2c} \\
  {\tilde G}_0^{-1} {\tilde G}_{D {\cal F}}
  = & {\tilde \Sigma}_{D {\cal F}} {\tilde G}_0 + \partial_{X^{\lambda}} {\cal F}_{\mu \nu}
  (\partial_{p_{\mu}} {\tilde G}_0^{-1} \partial_{p_{\lambda}} \partial_{p_{\nu}} {\tilde G}_0 - \partial_{p_{\lambda}} \partial_{p_{\mu}} {\tilde G}_0^{-1} \partial_{p_{\nu}} {\tilde G}_0)
  + 3 i (\partial_{p_{\lambda}} {\tilde G}_0^{-1} \partial_{X^{\lambda}} {\tilde G}_{\cal F} + \partial_{X^{\lambda}} {\tilde \Sigma}_{\cal F} \partial_{p_{\lambda}} {\tilde G}_0), \label{eq:dyson2d}
\end{align} \label{eq:dyson2}\end{subequations}
and from the right Dyson equation we obtain
\begin{subequations} \begin{align}
  {\tilde G}_0
  = & ({\tc L} - {\tilde \Sigma}_0)^{-1}, \label{eq:dyson3a} \\
  {\tilde G}_D {\tilde G}_0^{-1}
  = & {\tilde G}_0 {\tilde \Sigma}_D
  - i (\partial_{X^{\lambda}} {\tilde G}_0 \partial_{p_{\lambda}} {\tilde G}_0^{-1} - \partial_{p_{\lambda}} {\tilde G}_0 \partial_{X^{\lambda}} {\tilde G}_0^{-1}), \label{eq:dyson3b} \\
  {\tilde G}_{\cal F} {\tilde G}_0^{-1}
  = & {\tilde G}_0 {\tilde \Sigma}_{\cal F}
  - i {\cal F}_{\mu \nu} \partial_{p_{\mu}} {\tilde G}_0 \partial_{p_{\nu}} {\tilde G}_0^{-1}, \label{eq:dyson3c} \\
  {\tilde G}_{D {\cal F}} {\tilde G}_0^{-1}
  = & {\tilde G}_0 {\tilde \Sigma}_{D {\cal F}} + \partial_{X^{\lambda}} {\cal F}_{\mu \nu}
  (\partial_{p_{\mu}} {\tilde G}_0 \partial_{p_{\lambda}} \partial_{p_{\nu}} {\tilde G}_0^{-1} - \partial_{p_{\lambda}} \partial_{p_{\mu}} {\tilde G}_0 \partial_{p_{\nu}} {\tilde G}_0^{-1})
  - 3 i (\partial_{X^{\lambda}} {\tilde G}_{\cal F} \partial_{p_{\lambda}} {\tilde G}_0^{-1} + \partial_{p_{\lambda}} {\tilde G}_0 \partial_{X^{\lambda}} {\tilde \Sigma}_{\cal F}). \label{eq:dyson3d}
\end{align} \label{eq:dyson3}\end{subequations}
Note that $\partial_{X^{\lambda}}$ in Eqs.~\eqref{eq:dyson2d} and \eqref{eq:dyson3d} acts only on ${\cal F}_{\mu \nu}$ in ${\tilde G}_{\cal F}, {\tilde \Sigma}_{\cal F}$.
The left and right Dyson equations are equivalent and yield
\begin{subequations} \begin{align}
  {\tilde G}_D
  = & {\tilde G}_0 {\tilde \Sigma}_D {\tilde G}_0
  + i ({\tilde G}_0 (\partial_{X^{\lambda}} {\tilde G}_0^{-1} {\tilde G}_0 \partial_{p_{\lambda}} {\tilde G}_0^{-1}
  - \partial_{p_{\lambda}} {\tilde G}_0^{-1} {\tilde G}_0 \partial_{X^{\lambda}} {\tilde G}_0^{-1}) {\tilde G}_0, \label{eq:dyson4a} \\
  {\tilde G}_{\cal F}
  = & {\tilde G}_0 {\tilde \Sigma}_{\cal F} {\tilde G}_0
  + i {\cal F}_{\mu \nu} {\tilde G}_0 \partial_{p_{\mu}} {\tilde G}_0^{-1} {\tilde G}_0 \partial_{p_{\nu}} {\tilde G}_0^{-1} {\tilde G}_0, \label{eq:dyson4b} \\
  {\tilde G}_{D {\cal F}}
  = & {\tilde G}_0 {\tilde \Sigma}_{D {\cal F}} {\tilde G}_0
  + 3 i {\tilde G}_0 (\partial_{p_{\lambda}} {\tilde G}_0^{-1} {\tilde G}_0 \partial_{X^{\lambda}} {\tilde \Sigma}_{\cal F}
  - \partial_{X^{\lambda}} {\tilde \Sigma}_{\cal F} {\tilde G}_0 \partial_{p_{\lambda}} {\tilde G}_0^{-1}) {\tilde G}_0 \notag \\
  & - \partial_{X^{\lambda}} {\cal F}_{\mu \nu} {\tilde G}_0 (\partial_{p_{\mu}} {\tilde G}_0^{-1} {\tilde G}_0 \partial_{p_{\lambda}} \partial_{p_{\nu}} {\tilde G}_0^{-1}
  + \partial_{p_{\lambda}} \partial_{p_{\nu}} {\tilde G}_0^{-1} {\tilde G}_0 \partial_{p_{\mu}} {\tilde G}_0^{-1} \notag \\
  & + 2 \partial_{p_{\lambda}} {\tilde G}_0^{-1} {\tilde G}_0 \partial_{p_{\mu}} {\tilde G}_0^{-1} {\tilde G}_0 \partial_{p_{\nu}} {\tilde G}_0^{-1}
  + 2 \partial_{p_{\nu}} {\tilde G}_0^{-1} {\tilde G}_0 \partial_{p_{\mu}} {\tilde G}_0^{-1} {\tilde G}_0 \partial_{p_{\lambda}} {\tilde G}_0^{-1}) {\tilde G}_0. \label{eq:dyson4c}
\end{align} \label{eq:dyson4}\end{subequations}
Here we have used the Bianchi identity,
\begin{align}
  0
  = & (\partial_{X^{\lambda}} {\cal F}_{\mu \nu} + \partial_{X^{\mu}} {\cal F}_{\nu \lambda} + \partial_{X^{\nu}} {\cal F}_{\lambda \mu})
  {\tilde G}_0 \partial_{p_{\lambda}} {\tilde G}_0^{-1} {\tilde G}_0 \partial_{p_{\mu}} {\tilde G}_0^{-1} {\tilde G}_0 \partial_{p_{\nu}} {\tilde G}_0^{-1} {\tilde G}_0 \notag \\
  = & \partial_{X^{\lambda}} {\cal F}_{\mu \nu} {\tilde G}_0
  (\partial_{p_{\lambda}} {\tilde G}_0^{-1} {\tilde G}_0 \partial_{p_{\mu}} {\tilde G}_0^{-1} {\tilde G}_0 \partial_{p_{\nu}} {\tilde G}_0^{-1}
  + \partial_{p_{\mu}} {\tilde G}_0^{-1} {\tilde G}_0 \partial_{p_{\nu}} {\tilde G}_0^{-1} {\tilde G}_0 \partial_{p_{\lambda}} {\tilde G}_0^{-1}
  - \partial_{p_{\mu}} {\tilde G}_0^{-1} {\tilde G}_0 \partial_{p_{\lambda}} {\tilde G}_0^{-1} {\tilde G}_0 \partial_{p_{\nu}} {\tilde G}_0^{-1}) {\tilde G}_0. \label{eq:bianchi1}
\end{align}

The Keldysh Green's function contains three independent Green's functions; retarded, advanced, and lesser.
Among these, the lesser Green's function is necessary for calculating expectation values.
We express the lesser Green's function and self-energy as
\begin{subequations} \begin{align}
  G_{D {\cal F}}^<
  = & \pm [(G_{D {\cal F}}^{\rm R} - G_{D {\cal F}}^{\rm A}) f(-p_0) + G_{D {\cal F}}^{< (1)} f^{\prime}(-p_0) + G_{D {\cal F}}^{< (2)} f^{\prime \prime}(-p_0)], \label{eq:dyson5a} \\
  \Sigma_{D {\cal F}}^<
  = & \pm [(\Sigma_{D {\cal F}}^{\rm R} - \Sigma_{D {\cal F}}^{\rm A}) f(-p_0) + \Sigma_{D {\cal F}}^{< (1)} f^{\prime}(-p_0) + \Sigma_{D {\cal F}}^{< (2)} f^{\prime \prime}(-p_0)], \label{eq:dyson5b}
\end{align} \label{eq:dyson5}\end{subequations}
in which the upper or lower sign represents boson or fermion, and $f(\xi) = (e^{\beta \xi} \mp 1)^{-1}$ is the distribution function at temperature $T = \beta^{-1}$.
By substituting Eq.~\eqref{eq:dyson5} into Eq.~\eqref{eq:dyson4c}, we obtain
\begin{subequations} \begin{align}
  G_{D {\cal F}}^{\rm R}
  = & G_0^{\rm R} \Sigma_{D {\cal F}}^{\rm R} G_0^{\rm R}
  + 3 i G_0^{\rm R} [\partial_{p_{\lambda}} (G_0^{\rm R})^{-1} G_0^{\rm R} \partial_{X^{\lambda}} \Sigma_{\cal F}^{\rm R}
  - \partial_{X^{\lambda}} \Sigma_{\cal F}^{\rm R} G_0^{\rm R} \partial_{p_{\lambda}} (G_0^{\rm R})^{-1}] G_0^{\rm R} \notag \\
  & - \partial_{X^{\lambda}} {\cal F}_{\mu \nu} G_0^{\rm R} [\partial_{p_{\mu}} (G_0^{\rm R})^{-1} G_0^{\rm R} \partial_{p_{\lambda}} \partial_{p_{\nu}} (G_0^{\rm R})^{-1}
  + \partial_{p_{\lambda}} \partial_{p_{\nu}} (G_0^{\rm R})^{-1} G_0^{\rm R} \partial_{p_{\mu}} (G_0^{\rm R})^{-1} \notag \\
  & + 2 \partial_{p_{\lambda}} (G_0^{\rm R})^{-1} G_0^{\rm R} \partial_{p_{\mu}} (G_0^{\rm R})^{-1} G_0^{\rm R} \partial_{p_{\nu}} (G_0^{\rm R})^{-1}
  + 2 \partial_{p_{\nu}} (G_0^{\rm R})^{-1} G_0^{\rm R} \partial_{p_{\mu}} (G_0^{\rm R})^{-1} G_0^{\rm R} \partial_{p_{\lambda}} (G_0^{\rm R})^{-1}] G_0^{\rm R}, \label{eq:dyson6a} \\
  G_{D {\cal F}}^{\rm A}
  = & G_0^{\rm A} \Sigma_{D {\cal F}}^{\rm A} G_0^{\rm A}
  + 3 i G_0^{\rm A} [\partial_{p_{\lambda}} (G_0^{\rm A})^{-1} G_0^{\rm A} \partial_{X^{\lambda}} \Sigma_{\cal F}^{\rm A}
  - \partial_{X^{\lambda}} \Sigma_{\cal F}^{\rm A} G_0^{\rm A} \partial_{p_{\lambda}} (G_0^{\rm A})^{-1}] G_0^{\rm A} \notag \\
  & - \partial_{X^{\lambda}} {\cal F}_{\mu \nu} G_0^{\rm A} [\partial_{p_{\mu}} (G_0^{\rm A})^{-1} G_0^{\rm A} \partial_{p_{\lambda}} \partial_{p_{\nu}} (G_0^{\rm A})^{-1}
  + \partial_{p_{\lambda}} \partial_{p_{\nu}} (G_0^{\rm A})^{-1} G_0^{\rm A} \partial_{p_{\mu}} (G_0^{\rm A})^{-1} \notag \\
  & + 2 \partial_{p_{\lambda}} (G_0^{\rm A})^{-1} G_0^{\rm A} \partial_{p_{\mu}} (G_0^{\rm A})^{-1} G_0^{\rm A} \partial_{p_{\nu}} (G_0^{\rm A})^{-1}
  + 2 \partial_{p_{\nu}} (G_0^{\rm A})^{-1} G_0^{\rm A} \partial_{p_{\mu}} (G_0^{\rm A})^{-1} G_0^{\rm A} \partial_{p_{\lambda}} (G_0^{\rm A})^{-1}] G_0^{\rm A}, \label{eq:dyson6b} \\
  G_{D {\cal F}}^{< (1)}
  = & G_0^{\rm R} \Sigma_{D {\cal F}}^{< (1)} G_0^{\rm A}
  - 3 i [G_0^{\rm R} \partial_{X^0} \Sigma_{\cal F}^{\rm R} (G_0^{\rm R} - G_0^{\rm A}) - (G_0^{\rm R} - G_0^{\rm A}) \partial_{X^0} \Sigma_{\cal F}^{\rm A} G_0^{\rm A}] \notag \\
  & + 3 i G_0^{\rm R} [\partial_{p_{\lambda}} (G_0^{\rm R})^{-1} G_0^{\rm R} \partial_{X^{\lambda}} \Sigma_{\cal F}^{< (1)}
  - \partial_{X^{\lambda}} \Sigma_{\cal F}^{< (1)} G_0^{\rm A} \partial_{p_{\lambda}} (G_0^{\rm A})^{-1}] G_0^{\rm A} \notag \\
  & + \partial_{X^0} {\cal F}_{\mu \nu} \{G_0^{\rm R} \partial_{p_{\mu}} (G_0^{\rm R})^{-1} G_0^{\rm R} \partial_{p_{\nu}} [(G_0^{\rm R})^{-1} - (G_0^{\rm A})^{-1}] G_0^{\rm A}
  + G_0^{\rm R} \partial_{p_{\nu}} [(G_0^{\rm R})^{-1} - (G_0^{\rm A})^{-1}] G_0^{\rm A} \partial_{p_{\mu}} (G_0^{\rm A})^{-1} G_0^{\rm A} \notag \\
  & - 2 G_0^{\rm R} \partial_{p_{\nu}} (G_0^{\rm R})^{-1} G_0^{\rm R} \partial_{p_{\mu}} (G_0^{\rm R})^{-1} (G_0^{\rm R} - G_0^{\rm A})
  - 2 (G_0^{\rm R} - G_0^{\rm A}) \partial_{p_{\mu}} (G_0^{\rm A})^{-1} G_0^{\rm A} \partial_{p_{\nu}} (G_0^{\rm A})^{-1} G_0^{\rm A}\} \notag \\
  & + \partial_{X^{\lambda}} {\cal F}_{j0} \{G_0^{\rm R} \partial_{p_{\lambda}} \partial_{p_j} (G_0^{\rm R})^{-1} (G_0^{\rm R} - G_0^{\rm A})
  + (G_0^{\rm R} - G_0^{\rm A}) \partial_{p_{\lambda}} \partial_{p_j} (G_0^{\rm A})^{-1} G_0^{\rm A} \notag \\
  & + G_0^{\rm R} \partial_{p_j} (G_0^{\rm R})^{-1} G_0^{\rm R} \partial_{p_{\lambda}} [(G_0^{\rm R})^{-1} - (G_0^{\rm A})^{-1}] G_0^{\rm A}
  + G_0^{\rm R} \partial_{p_{\lambda}} [(G_0^{\rm R})^{-1} - (G_0^{\rm A})^{-1}] G_0^{\rm A} \partial_{p_j} (G_0^{\rm A})^{-1} G_0^{\rm A} \notag \\
  & - 2 G_0^{\rm R} \partial_{p_{\lambda}} (G_0^{\rm R})^{-1}  G_0^{\rm R} \partial_{p_j} (G_0^{\rm R})^{-1} (G_0^{\rm R} - G_0^{\rm A})
  - 2 (G_0^{\rm R} - G_0^{\rm A}) \partial_{p_j} (G_0^{\rm A})^{-1} G_0^{\rm A} \partial_{p_{\lambda}} (G_0^{\rm A})^{-1} G_0^{\rm A} \notag \\
  & + 2 G_0^{\rm R} \partial_{p_{\lambda}} (G_0^{\rm R})^{-1} (G_0^{\rm R} - G_0^{\rm A}) \partial_{p_j} (G_0^{\rm A})^{-1} G_0^{\rm A}
  + 2 G_0^{\rm R} \partial_{p_j} (G_0^{\rm R})^{-1} (G_0^{\rm R} - G_0^{\rm A}) \partial_{p_{\lambda}} (G_0^{\rm A})^{-1} G_0^{\rm A}\}, \label{eq:dyson6c} \\
  G_{D {\cal F}}^{< (2)}
  = & G_0^{\rm R} \Sigma_{D {\cal F}}^{< (2)} G_0^{\rm A}
  - \partial_{X^0} {\cal F}_{j0} \{G_0^{\rm R} \partial_{p_j} [(G_0^{\rm R})^{-1} - (G_0^{\rm A})^{-1}] G_0^{\rm A} + \partial_{p_j} (G_0^{\rm R} - G_0^{\rm A})\}. \label{eq:dyson6d}
\end{align} \label{eq:dyson6}\end{subequations}
$G_{D {\cal F}}^{< (1)}, G_{D {\cal F}}^{< (2)}$ are nonzero
only in the presence of dynamical electromagnetic fields $\partial_{X^0} {\cal F}_{\mu \nu}$ or a nonuniform electric field $\partial_{X^l} {\cal F}_{j0}$,
which drives the system into nonequilibrium.

\section{Derivation of Orbital Magnetic Quadrupole Moment} \label{sec:omq}
Next, we derive the orbital magnetic quadrupole moment by calculating the energy $K \equiv E - \mu N$ instead of the grand potential $\Omega \equiv E - T S - \mu N$.
The energy is expressed by
\begin{equation}
  K(X)
  = \pm i \hbar \int \frac{{\rm d}^D p}{(2 \pi \hbar)^D} (-p_0) \tr G^<, \label{eq:omq1}
\end{equation}
in which $D = d + 1$ is the spacetime dimension.
Therefore, the variation of the energy due to a nonuniform magnetic field is given by
\begin{equation}
  K_{D {\cal F}}(X)
  = \frac{i \hbar^3}{12} \int \frac{{\rm d}^D p}{(2 \pi \hbar)^D} f(-p_0) (-p_0) \tr (G_{D {\cal F}}^{\rm R}) + \cc, \label{eq:omq2}
\end{equation}
since the nonuniform magnetic field does not drive the system into nonequilibrium.
$G_{D {\cal F}}^{\rm R}$ has been already obtained in Eq.~\eqref{eq:dyson6a}.

In the absence of disorder or interactions, Eq.~\eqref{eq:dyson6a} is reduced to
\begin{align}
  g_{D {\cal F}}^{\rm R}
  = & -\partial_{X^{\lambda}} {\cal F}_{\mu \nu} g_0^{\rm R} [\partial_{p_{\mu}} (g_0^{\rm R})^{-1} g_0^{\rm R} \partial_{p_{\lambda}} \partial_{p_{\nu}} (g_0^{\rm R})^{-1}
  + \partial_{p_{\lambda}} \partial_{p_{\nu}} (g_0^{\rm R})^{-1} g_0^{\rm R} \partial_{p_{\mu}} (g_0^{\rm R})^{-1} \notag \\
  & + 2 \partial_{p_{\lambda}} (g_0^{\rm R})^{-1} g_0^{\rm R} \partial_{p_{\mu}} (g_0^{\rm R})^{-1} g_0^{\rm R} \partial_{p_{\nu}} (g_0^{\rm R})^{-1}
  + 2 \partial_{p_{\nu}} (g_0^{\rm R})^{-1} g_0^{\rm R} \partial_{p_{\mu}} (g_0^{\rm R})^{-1} g_0^{\rm R} \partial_{p_{\lambda}} (g_0^{\rm R})^{-1}] g_0^{\rm R} \notag \\
  = & -\partial_{X^{\lambda}} {\cal F}_{\mu \nu} \partial_{p_{\nu}} [g_0^{\rm R} \partial_{p_{\mu}} (g_0^{\rm R})^{-1} g_0^{\rm R} \partial_{p_{\lambda}} (g_0^{\rm R})^{-1} g_0^{\rm R}
  + g_0^{\rm R} \partial_{p_{\lambda}} (g_0^{\rm R})^{-1} g_0^{\rm R} \partial_{p_{\mu}} (g_0^{\rm R})^{-1} g_0^{\rm R}] \notag \\
  & - 2 \partial_{X^{\lambda}} {\cal F}_{\mu \nu}
  [g_0^{\rm R} \partial_{p_{\lambda}} (g_0^{\rm R})^{-1} g_0^{\rm R} \partial_{p_{\mu}} (g_0^{\rm R})^{-1} g_0^{\rm R} \partial_{p_{\nu}} (g_0^{\rm R})^{-1} g_0^{\rm R}
  + g_0^{\rm R} \partial_{p_{\nu}} (g_0^{\rm R})^{-1} g_0^{\rm R} \partial_{p_{\mu}} (g_0^{\rm R})^{-1} g_0^{\rm R} \partial_{p_{\lambda}} (g_0^{\rm R})^{-1} g_0^{\rm R}]. \label{eq:omq3}
\end{align}
$g_0^{\rm R}(\xi, {\vec p}) = [\xi - {\cal H}({\vec p}) + \mu + i \eta]^{-1}$ $(\eta \to +0)$ is the retarded Green's function of the Hamiltonian ${\cal H}({\vec p})$,
$\xi \equiv -p_0$,
and we have used the antisymmetric property of ${\cal F}_{\mu \nu}$.
By substituting Eq.~\eqref{eq:omq3} into Eq.~\eqref{eq:omq2}, we obtain
\begin{align}
  K_{D {\cal F}}
  = & -\frac{i \hbar^2}{6} \partial_{X^l} {\cal F}_{ij} \int \frac{{\rm d}^d p}{(2 \pi \hbar)^d} \int \frac{{\rm d} \xi}{2 \pi} f(\xi) \xi \notag \\
  & \times \tr [g_0^{\rm R} \partial_{p_l} (g_0^{\rm R})^{-1} g_0^{\rm R} \partial_{p_i} (g_0^{\rm R})^{-1} g_0^{\rm R} \partial_{p_j} (g_0^{\rm R})^{-1} g_0^{\rm R}
  + g_0^{\rm R} \partial_{p_j} (g_0^{\rm R})^{-1} g_0^{\rm R} \partial_{p_i} (g_0^{\rm R})^{-1} g_0^{\rm R} \partial_{p_l} (g_0^{\rm R})^{-1} g_0^{\rm R}] + \cc \label{eq:omq4}
\end{align}
This is Eq.~(9) in our Letter.
Note that the first term in Eq.~\eqref{eq:omq3} is the total derivative with respect to $p_j$ and can be dropped.
By expanding the trace with respect to the Bloch basis, we obtain
\begin{align}
  K_{D {\cal F}}
  = & \frac{i \hbar^2}{6} \partial_{X^l} {\cal F}_{ij} \sum_{nmr} \int \frac{{\rm d}^d p}{(2 \pi \hbar)^d}
  (\langle u_n | v^l | u_m \rangle \langle u_m | v^i | u_r \rangle \langle u_r | v^j | u_n \rangle + \cc)
  \int \frac{{\rm d} \xi}{2 \pi} f(\xi) \xi [(g_{0n}^{\rm R})^2 g_{0m}^{\rm R} g_{0r}^{\rm R} - \cc] \notag \\
  = & -\frac{i \hbar^2}{6} \partial_{X^l} {\cal F}_{ij} \sum_{n \not= m \not= r} \int \frac{{\rm d}^d p}{(2 \pi \hbar)^d}
  (\langle u_n | v^l | u_m \rangle \langle u_m | v^i | u_r \rangle \langle u_r | v^j | u_n \rangle + \cc) \notag \\
  & \times \int \frac{{\rm d} \xi}{2 \pi} f(\xi) \xi
  \left[\left(\frac{1}{\epsilon_n - \epsilon_m} + \frac{1}{\epsilon_n - \epsilon_r}\right) \frac{g_{0n}^{\rm R}}{(\epsilon_n - \epsilon_m) (\epsilon_n - \epsilon_r)}\right. \notag \\
  & \left.- \frac{g_{0m}^{\rm R}}{(\epsilon_m - \epsilon_n)^2 (\epsilon_m - \epsilon_r)} - \frac{g_{0r}^{\rm R}}{(\epsilon_r - \epsilon_n)^2 (\epsilon_r - \epsilon_m)}
  - \frac{(g_{0n}^{\rm R})^2}{(\epsilon_n - \epsilon_m) (\epsilon_n - \epsilon_r)} - \cc\right] \notag \\
  & - \frac{i \hbar^2}{6} \partial_{X^l} {\cal F}_{ij} \sum_{n \not= m} \int \frac{{\rm d}^d p}{(2 \pi \hbar)^d}
  (\langle u_n | v^l | u_m \rangle \langle u_m | v^i | u_m \rangle \langle u_m | v^j | u_n \rangle + \cc) \notag \\
  & \times \int \frac{{\rm d} \xi}{2 \pi} f(\xi) \xi
  \left[\frac{2 g_{0n}^{\rm R}}{(\epsilon_n - \epsilon_m)^3} + \frac{2 g_{0m}^{\rm R}}{(\epsilon_m - \epsilon_n)^3}
  - \frac{(g_{0n}^{\rm R})^2}{(\epsilon_n - \epsilon_m)^2} - \frac{(g_{0m}^{\rm R})^2}{(\epsilon_m - \epsilon_n)^2} - \cc\right] \notag \\
  & + \frac{i \hbar^2}{6} \partial_{X^l} {\cal F}_{ij} \sum_{n \not= m} \int \frac{{\rm d}^d p}{(2 \pi \hbar)^d}
  (\langle u_n | v^l | u_m \rangle \langle u_m | v^i | u_n \rangle \langle u_n | v^j | u_n \rangle + \cc) \notag \\
  & \times \int \frac{{\rm d} \xi}{2 \pi} f(\xi) \xi
  \left[\frac{g_{0n}^{\rm R}}{(\epsilon_n - \epsilon_m)^3} + \frac{g_{0m}^{\rm R}}{(\epsilon_m - \epsilon_n)^3}
  - \frac{(g_{0n}^{\rm R})^2}{(\epsilon_n - \epsilon_m)^2} + \frac{(g_{0n}^{\rm R})^3}{\epsilon_n - \epsilon_m} - \cc\right]. \label{eq:omq5}
\end{align}
$v^i \equiv \partial_{p_i} {\cal H}$ is the velocity operator, and $g_{0n}^{\rm R} = (\xi - \epsilon_n + \mu + i \eta)^{-1}$.
The other processes, $n = m \not= r$ and $n = m = r$, are forbidden by the antisymmetric property of ${\cal F}_{ij}$.
We carry out the integrals over $\xi$ to obtain
\begin{align}
  K_{D {\cal F}}
  = & -\frac{\hbar^2}{6} \partial_{X^l} {\cal F}_{ij} \sum_{n \not= m \not= r} \int \frac{{\rm d}^d p}{(2 \pi \hbar)^d}
  (\langle u_n | v^l | u_m \rangle \langle u_m | v^i | u_r \rangle \langle u_r | v^j | u_n \rangle + \cc) \notag \\
  & \times \left[\left(\frac{1}{\epsilon_n - \epsilon_m} + \frac{1}{\epsilon_n - \epsilon_r}\right) \frac{f_n (\epsilon_n - \mu)}{(\epsilon_n - \epsilon_m) (\epsilon_n - \epsilon_r)}\right. \notag \\
  & \left.- \frac{f_m (\epsilon_m - \mu)}{(\epsilon_m - \epsilon_n)^2 (\epsilon_m - \epsilon_r)} - \frac{f_r (\epsilon_r - \mu)}{(\epsilon_r - \epsilon_n)^2 (\epsilon_r - \epsilon_m)}
  - \frac{f_n + f_n^{\prime} (\epsilon_n - \mu)}{(\epsilon_n - \epsilon_m) (\epsilon_n - \epsilon_r)}\right] \notag \\
  & - \frac{\hbar^2}{6} \partial_{X^l} {\cal F}_{ij} \sum_{n \not= m} \int \frac{{\rm d}^d p}{(2 \pi \hbar)^d}
  (\langle u_n | v^l | u_m \rangle \langle u_m | v^i | u_m \rangle \langle u_m | v^j | u_n \rangle + \cc) \notag \\
  & \times \left[\frac{2 f_n (\epsilon_n - \mu)}{(\epsilon_n - \epsilon_m)^3} + \frac{2 f_m (\epsilon_m - \mu)}{(\epsilon_m - \epsilon_n)^3}
  - \frac{f_n + f_n^{\prime} (\epsilon_n - \mu)}{(\epsilon_n - \epsilon_m)^2} - \frac{f_m + f_m^{\prime} (\epsilon_m - \mu)}{(\epsilon_m - \epsilon_n)^2}\right] \notag \\
  & + \frac{\hbar^2}{6} \partial_{X^l} {\cal F}_{ij} \sum_{n \not= m} \int \frac{{\rm d}^d p}{(2 \pi \hbar)^d}
  (\langle u_n | v^l | u_m \rangle \langle u_m | v^i | u_n \rangle \langle u_n | v^j | u_n \rangle + \cc) \notag \\
  & \times \left[\frac{f_n (\epsilon_n - \mu)}{(\epsilon_n - \epsilon_m)^3} + \frac{f_m (\epsilon_m - \mu)}{(\epsilon_m - \epsilon_n)^3}
  - \frac{f_n + f_n^{\prime} (\epsilon_n - \mu)}{(\epsilon_n - \epsilon_m)^2} + \frac{1}{2} \frac{2 f_n^{\prime} + f_n^{\prime \prime} (\epsilon_n - \mu)}{\epsilon_n - \epsilon_m}\right] \notag \\
  = & -\frac{\hbar^2}{6} \sum_{n \not= m \not= r} \int \frac{{\rm d}^d p}{(2 \pi \hbar)^d}
  \frac{\langle u_n | v^l | u_m \rangle \langle u_m | v^i | u_r \rangle \langle u_r | v^j | u_n \rangle + \cc}{(\epsilon_n - \epsilon_m)^2 (\epsilon_n - \epsilon_r)} \notag \\
  & \times \{(2 \partial_{X^l} {\cal F}_{ij} - \partial_{X^i} {\cal F}_{jl} - \partial_{X^j} {\cal F}_{li}) f_n (\epsilon_n - \mu)
  - \partial_{X^l} {\cal F}_{ij} (\epsilon_n - \epsilon_m) [f_n + f_n^{\prime} (\epsilon_n - \mu)]\} \notag \\
  & - \frac{\hbar^2}{6} \partial_{X^l} {\cal F}_{ij} \sum_{n \not= m} \int \frac{{\rm d}^d p}{(2 \pi \hbar)^d}
  \frac{\langle u_n | v^l | u_m \rangle \langle u_m | v^i | u_m \rangle \langle u_m | v^j | u_n \rangle + \cc}{(\epsilon_n - \epsilon_m)^3}
  \{3 f_n (\epsilon_n - \mu) - (\epsilon_n - \epsilon_m) [f_n + f_n^{\prime} (\epsilon_n - \mu)]\} \notag \\
  & - \frac{\hbar^2}{6} \partial_{X^l} {\cal F}_{ij} \sum_{n \not= m} \int \frac{{\rm d}^d p}{(2 \pi \hbar)^d}
  \frac{\langle u_n | v^l | u_m \rangle \langle u_m | v^j | u_n \rangle \langle u_n | v^i | u_n \rangle + \cc}{(\epsilon_n - \epsilon_m)^3} \notag \\
  & \times \{3 f_n (\epsilon_n - \mu) - 2 (\epsilon_n - \epsilon_m) [f_n + f_n^{\prime} (\epsilon_n - \mu)]
  + (\epsilon_n - \epsilon_m)^2 [2 f_n^{\prime} + f_n^{\prime \prime} (\epsilon_n - \mu)]/2\}. \label{eq:omq6}
\end{align}
$f_n \equiv f(\epsilon_n - \mu)$ is the distribution function.
In the first term, we use the Bianchi identity again and obtain
\begin{subequations} \begin{align}
  {\tilde M}^l_{\phantom{l} k}
  = & \frac{q}{\hbar} \frac{1}{2} \epsilon_{ijk} \sum_n \int \frac{{\rm d}^d p}{(2 \pi \hbar)^d}
  \{A_n^{lij} f_n (\epsilon_n - \mu) + m_n^{lij} [f_n + f_n^{\prime} (\epsilon_n - \mu)] + \gamma_n^{lij} [2 f_n^{\prime} + f_n^{\prime \prime} (\epsilon_n - \mu)]\}, \label{eq:omq7a} \\
  A_n^{lij}
  \equiv & \frac{\hbar^3}{2} \sum_{m, r (\not= n)} \frac{\langle u_n | v^l | u_m \rangle \langle u_m | v^i | u_r \rangle \langle u_r | v^j | u_n \rangle}{(\epsilon_n - \epsilon_m)^2 (\epsilon_n - \epsilon_r)}
  + \frac{\hbar^3}{2} \sum_{m (\not= n)} \frac{\langle u_n | v^l | u_m \rangle \langle u_m | v^j | u_n \rangle \langle u_n | v^i | u_n \rangle}{(\epsilon_n - \epsilon_m)^3}
  + \cc - (i \leftrightarrow j) \notag \\
  = & \frac{\hbar^3}{2} \sum_{m (\not= n)} \frac{\langle \partial_{p_l} u_n | u_m \rangle \langle u_m | \partial_{p_i} (\epsilon_n + {\cal H}) Q_n | \partial_{p_j} u_n \rangle}{\epsilon_n - \epsilon_m}
  + \cc - (i \leftrightarrow j), \label{eq:omq7b} \\
  m_n^{lij}
  \equiv & -\frac{\hbar^3}{6} \sum_{m, r (\not= n)} \frac{\langle u_n | v^l | u_m \rangle \langle u_m | v^i | u_r \rangle \langle u_r | v^j | u_n \rangle}{(\epsilon_n - \epsilon_m) (\epsilon_n - \epsilon_r)}
  - \frac{\hbar^3}{3} \sum_{m (\not= n)} \frac{\langle u_n | v^l | u_m \rangle \langle u_m | v^j | u_n \rangle \langle u_n | v^i | u_n \rangle}{(\epsilon_n - \epsilon_m)^2}
  + \cc - (i \leftrightarrow j) \notag \\
  = & -\frac{\hbar^3}{6} \langle \partial_{p_l} u_n | Q_n \partial_{p_i} (2 \epsilon_n + {\cal H}) Q_n | \partial_{p_j} u_n \rangle
  + \cc - (i \leftrightarrow j), \label{eq:omq7c} \\
  \gamma_n^{lij}
  \equiv & \frac{\hbar^3}{12} \sum_{m (\not= n)} \frac{\langle u_n | v^l | u_m \rangle \langle u_m | v^j | u_n \rangle \langle u_n | v^i | u_n \rangle}{\epsilon_n - \epsilon_m}
  + \cc - (i \leftrightarrow j) \notag \\
  = & \frac{\hbar^3}{12} \langle \partial_{p_l} u_n | (\epsilon_n - {\cal H}) | \partial_{p_j} u_n \rangle \partial_{p_i} \epsilon_n
  + \cc - (i \leftrightarrow j). \label{eq:omq7d}
\end{align} \label{eq:omq7}\end{subequations}
This is Eq.~(10) in our Letter.
Here we have used $\langle u_m | v^j | u_n \rangle = (\epsilon_n - \epsilon_m) \langle u_m | \partial_{p_j} u_n \rangle$ for $m \not= n$,
and $Q_n \equiv 1 - | u_n \rangle \langle u_n |$ appears from the summation over $m, r \not= n$.
For degenerate bands, we only have to replace the projection operator with $Q_n \equiv 1 - \sum_s | u_{ns} \rangle \langle u_{ns} |$,
in which $s$ indicates the index of degenerate bands with energy $\epsilon_n$.
There is no additional contribution because the interband processes between degenerate bands are forbidden by $\langle u_{ns} | v^j | u_{nt} \rangle = 0$ for $s \not= t$.
Eq.~(6) in our Letter can be solved using
\begin{subequations} \begin{align}
  f(\xi) \xi
  = & \frac{\partial}{\partial \beta} \left\{\beta \left[-\int_{\xi}^{\infty} {\rm d} z f(z)\right]\right\}
  = \frac{\partial}{\partial \beta} \{\beta [\pm \beta^{-1} \ln (1 \mp e^{-\beta \xi})]\}, \label{eq:omq8a} \\
  f(\xi) + f^{\prime}(\xi) \xi
  = & \frac{\partial}{\partial \beta} [\beta f(\xi)], \label{eq:omq8b} \\
  2 f^{\prime}(\xi) + f^{\prime \prime}(\xi) \xi
  = & \frac{\partial}{\partial \beta} [\beta f^{\prime}(\xi)]. \label{eq:omq8c}  
\end{align} \label{eq:omq8}\end{subequations}